# The Baryon Catastrophe and the multiphase intracluster medium

Katherine F. Gunn* and Peter A. Thomas
*Astronomy Centre, School of Mathematical and Physical Sciences, University of Sussex, Falmer, Brighton BN1 9QH*

**ABSTRACT**
In this paper, we review the theories and observations which together have led to the concept of the Baryon Catastrophe: observations of the baryon fraction on the scale of clusters of galaxies appear to be at least three times as high as the universal baryon fraction predicted by the theory of primordial nucleosynthesis in a flat, $\Omega_0 = 1$, universe. We have investigated whether this discrepancy could be eliminated by treating the intracluster gas as a multiphase medium, and found that this treatment both lowers the calculated mass of gas in a cluster and increases the inferred gravitational potential. These combined effects can reduce the calculated baryon fraction by between a quarter and a half: the precise amount depends upon the volume fraction distribution of density phases in the gas but is independent of the temperature profile across the cluster. Thus moving to a multiphase intracluster medium cannot resolve the Baryon Catastrophe by itself; other possible causes and explanations are discussed.

**Key words:**  cooling flows — dark matter

## 1 INTRODUCTION

The concept of the Baryon Catastrophe was first formulated when it was realized that clusters of galaxies contain more than their fair share of baryonic matter. Primordial nucleosynthesis is one of the most well understood theories of modern physics and predicts that the density of baryonic matter in units of the critical closure density, $\Omega_b$, lies in the range, $0.010 < \Omega_b h^2 < 0.015$ (Walker et al. 1991), where $H_0 = 100 h$ km s$^{-1}$ Mpc$^{-1}$. The theory of inflation predicts that the density of the universe is very close to the critical density, $\Omega = 1$, and this is supported by observations of peculiar velocity fields and large scale structure (see e.g. Dekel 1994). If these two theories are to be compatible with one another, then non-baryonic matter is needed in addition to the baryonic matter, with a density of $\sim 100 - 1.25 h^{-2}$ percent of the critical density. Even if a very low value of the Hubble parameter is assumed, such as $h = 0.3$, the Universe should still contain only 14 percent baryonic matter compared with 86 percent non-baryonic matter, giving a baryon fraction of 0.16.

The largest scale on which the baryon fraction can be reliably calculated is that of clusters of galaxies. Recent observations (Fabian 1991; White et al. 1993; Böhringer 1994; Henriksen & Mamon 1994; Allen et al. 1995; Buote & Canizares 1995; Elbaz, Arnaud & Böhringer 1995; White & Fabian 1995) have found that clusters of galaxies contain three times as many baryons as predicted by the theory of primordial nucleosynthesis in a critical density universe.

We are left with the following options: (i) the observations, or their interpretation, are incorrect, (ii) $\Omega_0 < 1$, or (iii) the limits from primordial nucleosynthesis must be relaxed (see e.g. Sasselov & Goldwirth 1995). This paper tests the first of these possibilities. We show that treating the intracluster medium (icm) as multiphase (in which there is an emulsion of gas blobs of differing density at any given radius) can lower the estimate of the baryon fraction by as much as 26 to 49 percent. The following section provides the motivation for and a formal description of the multiphase icm. Our cluster model is described in Section 3 and the results presented in Section 4. The implications for, and possible resolutions of, the Baryon Catastrophe are discussed in the final section.

## 2 THE MULTIPHASE INTRACLUSTER MEDIUM

### 2.1 Motivation: cooling flows in clusters of galaxies

The treatment of the intracluster gas as a multiphase medium is motivated by the study of cooling flows in clusters of galaxies. The central gas density in clusters is extremely high, therefore close encounters between particles

---

* Current address: Dept. of Physics, University of Durham, South Road, Durham, DH1 3LE

are frequent, and energy is lost due to the emission of X-rays by thermal bremsstrahlung and, at lower temperatures, by line radiation. The cooling process is thermally unstable and overdense blobs of gas will rapidly cool further, condensing into molecular clouds or low-mass stars. As matter in the centre of the cluster cools, more matter moves inwards to maintain pressure support, so forming a cooling flow. For a review of cooling flows see Fabian (1994).

The amount of gas deposited in these cooling flows is considerable, up to 1000 $M_\odot$ yr$^{-1}$, (Thomas, Fabian & Nulsen 1987), and may contribute to the formation of a central dominant galaxy. The matter is deposited not solely in the centre of the cluster, however, but over a wide range of radii, out to at least $150 h^{-1}$ kpc. In order for this to happen, there must be a range of densities and temperatures at each radius in the cluster. If this were not the case and, as is usually assumed in cluster models, there were a single density and temperature corresponding to each radius, then the gas would cool catastrophically at the centre of the cluster, giving a much more centrally-condensed surface brightness profile in X-rays than is observed.

The cooling flow models only constrain the structure of the icm within the region in which the cooling time is much less than the age of the cluster. There is no reason to suppose, however, that a single-phase model is appropriate at larger radii. The history of the icm is poorly understood but is likely to be very complex. In particular, the moderate metallicity (approximately half solar) suggests that it has been lost from galaxies either by stripping or by galactic winds. In addition, mergers of small groups and subclusters may introduce and mix together gas of a wide range of entropies. If the gas phases were free to move past one another then convection would stratify them on a dynamical time. However, the cooling flow observations show that this does not happen, at least in the cluster core. There are a variety of physical processes which may prevent convection such as magnetic pinning, or simply viscous drag (see the discussion in Nulsen 1986). Here we assume that the icm is multiphase throughout in order to see what effect this has on estimates of the baryon fraction.

### 2.2 Description: the volume fraction

The formalism for describing a multiphase icm was set out by Nulsen (1986). He introduced the volume fraction distribution, $f(\rho, r, t)$, such that $f \mathrm{d}f$ denotes the fraction of the volume at radius $r$ and time $t$ which contains gas in the density range $[\rho, \rho+\mathrm{d}\rho]$. Under the assumption that the density phases at any particular radius are in pressure equilibrium (which will be a good approximation until the cooling time drops below the sound-crossing time of the individual gas blobs) and that they are comoving, then it is possible to derive multiphase versions of the fluid equations which describe both the motion of gas through the cluster and the time-evolution of the volume fraction. Throughout the bulk of the cluster, where the mean cooling time is much longer than the age of the cluster, the distribution of $f$ will change very slowly and, in the absence of a detailed model for the formation of the icm, is poorly constrained. Within the cooling flow, however, $f$ is strongly modulated by cooling such that, at high densities,

$$f(\rho) \sim \rho^{-(4-\alpha)}, \qquad (1)$$

where $\alpha$ is the slope of the cooling function (see below).

A heuristic argument for the above form of the distribution function is as follows. Consider an initial distribution, $f_i$. If the phases of material at the highest densities of $f_i$ are considered, then their cooling time is relatively short. If this timescale is less than a Hubble time, then these phases will already have cooled, and will not feature in the volume fraction distribution function today. On the other hand, the phases of material at the lowest densities will have a cooling time which exceeds the Hubble time, and will not have cooled by the present. We are therefore left with a very narrow range of initial densities which are in the process of cooling today, for which the value of $f_i$ is approximately constant. The functional form of the highest density phases will therefore be approximately independent of the initial distribution.

The evolution of $f$ is governed by the Energy Equation

$$T \frac{\mathrm{d}s}{\mathrm{d}t} = -\frac{\xi}{\rho}, \qquad (2)$$

where $T$ is the temperature, $s$ the specific entropy, and $\xi$ is the emissivity (i.e. the emission per unit volume). For an Equation of State

$$P = \frac{2}{3} u, \qquad (3)$$

where $P$ is the pressure and $u$ the specific thermal energy of the gas, then

$$s = \frac{3}{2} \frac{k}{\mu m_H} \ln\left(\frac{P}{\rho^{5/3}}\right), \qquad (4)$$

where $k$ is the Boltzmann constant and $\mu m_H$ the mass per particle. Writing

$$\xi = n^2 \Lambda(T) = n^2 \Lambda_0 T^\alpha, \qquad (5)$$

where $n = \rho/\mu m_H$ and $\alpha = \frac{1}{2}$ for thermal bremsstrahlung, then a little rearrangement gives

$$\frac{\dot{\rho}}{\rho} = \frac{3}{5} \frac{\dot{P}}{P} + \frac{2}{5} \frac{n^2 \Lambda}{P}. \qquad (6)$$

At high densities the second term on the right-hand-side of the above equation dominates. Then

$$\dot{\rho} \sim \frac{2}{5} \frac{\rho n^2 \Lambda}{P} \propto \rho^{3-\alpha}, \qquad (7)$$

where we have used the fact that the phases are in pressure equilibrium, $\rho T =$ constant.

The equation of mass conservation in spatial and density space is

$$\frac{\partial}{\partial t}(\rho f) + \nabla \cdot (\rho f \vec{v}) + \frac{\partial}{\partial \rho}(\rho f \dot{\rho}) = 0. \qquad (8)$$

Unless the functional form of $f$ changes very rapidly (which we have argued above is very unlikely) then the final term dominates at high density, giving

$$\rho f \dot{\rho} \sim \text{constant}, \qquad (9)$$

or $f \sim \rho^{-(4-\alpha)}$ as in Equation 1.

The above argument is only suggestive but was given some credence by the simulations of Thomas (1988) who followed numerically the time-evolution of a range of volume fractions. In every case a high-density tail of the correct form was reproduced. In fact, all volume fractions, $f_i$, which were initially restricted to a narrow range of densities, rapidly converged towards a pure power law,

$$f_1 = \begin{cases} (3-\alpha)\rho_{\min}^{3-\alpha}\rho^{-(4-\alpha)} & \rho > \rho_{\min}, \\ 0 & \rho < \rho_{\min}, \end{cases} \quad (10)$$

where the constant of proportionality has been chosen so as to give a total volume fraction of unity when integrated over all densities. A multiphase, constant-pressure cooling flow model with this volume fraction distribution gives a mass-deposition profile $\dot{M} \propto r^{9/8}$ which is conveniently close to that expected for the distribution of dark matter around the central dominant galaxies in clusters. This suggests that $f_1$ may be ubiquitous in nature and may be taken as a canonical form for the volume fraction. However $f_1$ is just one of a whole family of solutions to the steady-state cooling flow equations derived in Nulsen (1986). It lies at one extreme with the following distribution at the other:

$$f_2 = \frac{2-\alpha}{\Gamma\left(\frac{3-\alpha}{2-\alpha}\right)}\rho_0^{3-\alpha}e^{-(\rho_0/\rho)^{2-\alpha}}. \quad (11)$$

$f_2$ gives a much more extended mass-deposition profile, $\dot{M} \propto r^3$, at variance with the observations, but it is difficult to rule it out on purely theoretical grounds (it is possible that the comoving assumption may break down when gas phases of arbitrarily low density are included in the emulsion).

In this paper we use $f_1$ and $f_2$ to represent the extremes of the possible multiphase behaviour found in clusters. We emphasize once again that the cooling time in the bulk of a cluster is very long, so that one would not expect the steady-state cooling flow equations to give an accurate representation of the system. Nevertheless, the above distributions provide a reasonable estimate of the expected range of behaviour.

## 3 METHOD

### 3.1 The cluster model

We will suppose that the distributions of gas and mass in the cluster are determined via X-ray observations. The emissivity profile is often parameterized as

$$\xi(r) = \xi_0 \left(1 + \frac{r^2}{a^2}\right)^{-3\beta}, \quad (12)$$

where the core radius, $a$, and slope, $\beta$, can be determined by fitting to the projected surface brightness profile. Jones & Forman (1984) give typical values $a \lesssim 200h^{-1}$ kpc and $\beta \approx \frac{2}{3}$. Direct deprojection of the surface brightness profiles (as described by Fabian et al. 1981) gives very similar results. Thus the gas is much more extended than the galaxy distribution which declines as $r^{-3}$ in the outskirts of the cluster.

Our first constraint on the system is therefore

$$\Lambda_0 < n^2 T^\alpha > = \xi_0 \left(1 + \frac{r^2}{a^2}\right)^{-3\beta}, \quad (13)$$

where $<>$ represents an average over all density phases,

$$< A > \equiv \int A f \mathrm{d}\rho. \quad (14)$$

A second constraint is required in order to determine the density and temperature at all radii. In principle this could be provided by a direct measurement of the temperature profile, but unfortunately the observations are generally inadequate to provide more than an overall spectrum for the whole cluster (spatially-resolved spectra are now becoming available from ASCA but even these are sufficient only to indicate the sense of the temperature variation, whether approximately isothermal or declining with radius). We therefore assume a polytropic relation,

$$P \propto < \rho >^\gamma, \quad (15)$$

where $1 \leq \gamma \leq 5/3$. The lower limit corresponds to an isothermal cluster and the upper to one which is on the verge of convective instability. We ignore the effect of cooling flows which may give a temperature decline within the core of the cluster.

The emission-weighted temperature averaged over the whole cluster is assumed to be determined by the X-ray observations,

$$T_X = \frac{\int < n^2 T^{1+\alpha} > 4\pi r^2 \mathrm{d}r}{\int < n^2 T^\alpha > 4\pi r^2 \mathrm{d}r}. \quad (16)$$

For the purposes of illustration we also define an emission-weighted temperature as a function of radius,

$$< T >_{\mathrm{ew}} = \frac{< n^2 T^{1+\alpha} >}{< n^2 T^\alpha >}. \quad (17)$$

The distribution of gas in the cluster can be related to the total mass via the Equation of Hydrostatic Support which, for a spherically-symmetric system, has the following form,

$$\frac{\mathrm{d}P}{\mathrm{d}r} = - < \rho > \frac{GM_{\mathrm{tot}}}{r^2}, \quad (18)$$

where $G$ is the Gravitational constant and $M_{\mathrm{tot}}(r)$ is the total mass interior to radius $r$.

Equations 13, 15, 16 and 18 together determine the distribution of both the gas and the mass within the cluster. The single-phase equations are obtained by merely omitting the average over density phases. Before discussing the solutions, we will first convert the equations into dimensionless form.

### 3.2 The dimensionless equations

We define dimensionless variables (denoted by primes) as follows:

$$\begin{aligned} r &= ar' \\ n &= \eta n' \\ T &= T_X T' \\ P &= \eta k T_X P' \\ M_{\mathrm{gas}} &= \mu_{\mathrm{gas}} M'_{\mathrm{gas}} \\ M_{\mathrm{tot}} &= \mu_{\mathrm{tot}} M'_{\mathrm{tot}}. \end{aligned} \quad (19)$$

With these definitions, Equation 16 becomes

$$\int < n'^2 T'^{1+\alpha} > r'^2 \mathrm{d}r' = \int < n'^2 T'^\alpha > r'^2 \mathrm{d}r'. \quad (20)$$

Choosing $\eta$ to satisfy $\xi_0 = \Lambda_0 \eta^2 T_X^\alpha$ gives

$$< n'^2 T'^\alpha > = (1 + r'^2)^{-3\beta}. \quad (21)$$

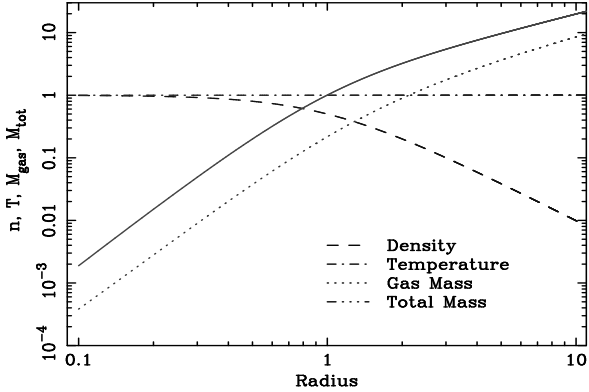

**Figure 1.** The temperature, density and cumulative gas and total mass profiles for a single-phase, isothermal cluster.

Finally, we set $\mu_{\rm gas} = 4\pi a^3 \mu m_H \eta$ and $\mu_{\rm tot} = akT_X/G\mu m_H$, giving

$$M'_{\rm gas} = \int <n'> r'^2 {\rm d}r' \qquad (22)$$

and

$$M'_{\rm tot} = -\frac{r'^2}{<n'>}\frac{{\rm d}P'}{{\rm d}r'}. \qquad (23)$$

The relative normalization of $M_{\rm gas}$ and $M_{\rm tot}$ is dependent upon $\xi_0$, $\Lambda_0$ and a whole host of observational parameters. Here we are concerned only with the difference between the normalization under the single-phase and multiphase assumptions.

For clarity in the discussion that follows we will henceforce drop the primes in the above equations. Thus all variables are to be considered dimensionless.

## 4 RESULTS

### 4.1 The isothermal case

We start with the simpler case in which the icm is assumed to be isothermal (i.e. there is the same emission-weighted temperature at each radius). Figure 1 shows the variation of density, temperature, gas mass and total mass with radius for a single-phase icm. In this and the following figure we fix $\alpha = \frac{1}{2}$ and $\beta = \frac{2}{3}$: their precise values make little difference to our conclusions. We stress once again that the relative normalization of the two mass curves in our model is arbitrary. Thus, although the baryon fraction can be seen to increase as one moves out through the cluster, its magnitude cannot be determined from this plot.

Now consider the multiphase case. As temperature is independent of radius, we have from Equation 20

$$<n^2T^{1+\alpha}> = <n^2T^\alpha> \qquad (24)$$

(recall that we are omitting the primes for clarity of expression). Hence

$$P<n^{1-\alpha}> = <n^{2-\alpha}> \qquad (25)$$

or

$$P = <n> C_T \qquad (26)$$

where

**Table 1.** Scaling factors for the gas mass, total mass and baryon fraction, relative to the single-phase case, for the multiphase volume fraction distribution $f_1$.

| $\alpha$ | $C_n$ | $C_T$ | $C_M$ |
|---|---|---|---|
| | $\dfrac{(3-\alpha)^{1/2}}{2^{\alpha/2}(2-\alpha)}$ | $\dfrac{2(2-\alpha)}{(3-\alpha)}$ | $\dfrac{(3-\alpha)^{3/2}}{2^{(2+\alpha)/2}(2-\alpha)^2}$ |
| 0 | 0.87 | 1.33 | 0.65 |
| $\frac{1}{2}$ | 0.89 | 1.20 | 0.74 |

**Table 2.** Scaling factors for the gas mass, total mass and baryon fraction, relative to the single-phase case, for the multiphase volume fraction distribution $f_2$.

| $\alpha$ | $C_n$ | $C_T$ | $C_M$ |
|---|---|---|---|
| | $\dfrac{(2-\alpha)^{\frac{1}{2}}\Gamma\left(\frac{2}{2-\alpha}\right)^{\frac{\alpha}{2}}}{\Gamma\left(\frac{1}{2-\alpha}\right)^{\frac{2+\alpha}{2}}}$ | $\dfrac{\Gamma\left(\frac{1}{2-\alpha}\right)^2}{(2-\alpha)\Gamma\left(\frac{2}{2-\alpha}\right)}$ | $\dfrac{(2-\alpha)^{\frac{3}{2}}\Gamma\left(\frac{2}{2-\alpha}\right)^{\frac{2+\alpha}{2}}}{\Gamma\left(\frac{1}{2-\alpha}\right)^{\frac{6+\alpha}{2}}}$ |
| 0 | 0.80 | 1.57 | 0.51 |
| $\frac{1}{2}$ | 0.82 | 1.37 | 0.60 |

$$C_T \equiv \frac{<n^{2-\alpha}>}{<n><n^{1-\alpha}>} \qquad (27)$$

is a dimensionless constant which depends upon the functional form of the volume fraction distribution.

From Equation 21, we see that the emissivity depends upon

$$<n^2T^\alpha> = P^\alpha <n^{2-\alpha}> = <n>^2 \frac{<n^{2-\alpha}>}{<n>^{2-\alpha}} C_T^\alpha. \qquad (28)$$

Hence

$$<n> = C_n(1+r^2)^{-3\beta/2} \qquad (29)$$

where

$$C_n \equiv \left(\frac{<n^{2-\alpha}>}{<n^{2-\alpha}> C_T^\alpha}\right)^{\frac{1}{2}} \equiv \frac{<n><n^{1-\alpha}>^{\alpha/2}}{<n^{2-\alpha}>^{(1+\alpha)/2}}. \qquad (30)$$

The multiphase solution is therefore just a simple scaling of the single-phase one with the gas density being reduced by a factor $C_n$. Similarly, from Equations 23 and 26, it is easy to see that the total mass is scaled upwards by a factor $C_T$ relative to the single-phase case. The baryon fraction is therefore lowered by a factor

$$C_M \equiv \frac{C_n}{C_T} \equiv \frac{<n>^2<n^{1-\alpha}>^{(2+\alpha)/2}}{<n^{2-\alpha}>^{(3+\alpha)/2}}. \qquad (31)$$

Tables 1 and 2 give general expressions for the form-factors $C_n$, $C_T$ and $C_M$, plus numerical values for two different values of $\alpha$, for the two volume fraction distributions discussed in Section 2.2. The canonical form, $f_1$, offers a reduction of 26 percent to the inferred baryon fraction at high temperatures, rising to 35 percent in lower temperature clusters for which a lower value of $\alpha$ is more appropriate. The second form of the volume fraction, $f_2$, that has a higher proportion of low-density phases, gives an even greater reduction of between 40 and 49 percent.

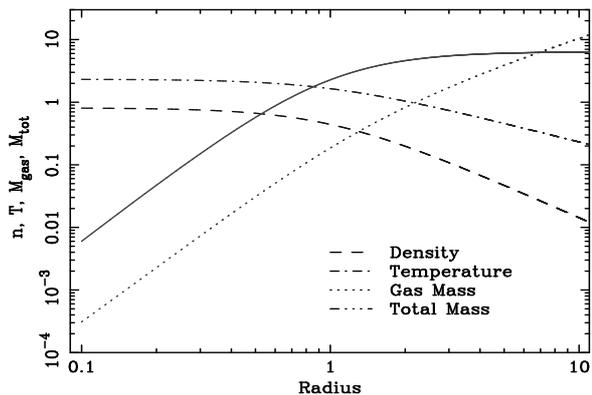

**Figure 2.** The temperature, density and cumulative gas and total mass profiles for a single-phase, polytropic cluster with $\gamma = 11/7$.

### 4.2 The polytropic case

Next we investigate a more general temperature distribution of the form given by Equation 3. More specifically, we write

$$P = \kappa <n>^\gamma \qquad (32)$$

where $\kappa$ is constant across the cluster. Increasing $\gamma$ results in a higher central temperature and hence a higher inferred gravitational potential. Thus the baryon fraction is lowered in the cluster core. At large radii, however, the reverse is true. Indeed, the condition

$$\gamma - 1 \leq \frac{2}{6\beta - \alpha} \qquad (33)$$

must hold if the enclosed mass is not to decline (unphysically) at large radii. Figure 2 shows the equivalent profiles to Figure 1 for the limiting case $\gamma = 11/7 \approx 1.57$ ($\alpha = \frac{1}{2}$, $\beta = \frac{2}{3}$). The central baryon fraction is just one quarter of that in the isothermal case, but by $r = 3$ it is approximately equal and at larger radii it increases rapidly. Moving to a polytropic temperature law does not, therefore, help to alleviate the baryon catastrophe.

Although the formal derivation is harder than for the isothermal case, the form factors which correct the estimates of the gas and total mass are identical. This is because the ratio of the 'dynamical' temperature, $P/<n>$, to the emission-weighted temperature depends only on the volume fraction distribution and not on the variation of $P$ and $<n>$ across the cluster.

To derive this result, consider first Equation 21. The left-hand-side can be written as

$$\begin{aligned}<n^2 T^\alpha> &= P^\alpha <n^{2-\alpha}> \\ &= \kappa^\alpha \frac{<n^{2-\alpha}>}{<n>^{2-\alpha}} <n>^{2+\alpha(\gamma-1)} \; . \end{aligned} \qquad (34)$$

Hence

$$<n> = A(1 + r^2)^{-3\beta/(2+\alpha(\gamma-1))} \qquad (35)$$

where

$$A = \left( \frac{<n>^{2-\alpha}}{<n^{2-\alpha}> \kappa^\alpha} \right)^{\frac{1}{2+\alpha(\gamma-1)}} . \qquad (36)$$

Contrast this with Equations 29 and 30 for the isothermal case.

Equation 35 can now be substituted into the temperature constraint, Equation 20, to yield after some manipulation

$$\kappa A^{\gamma-1} = C_T I_\gamma \qquad (37)$$

where $C_T$ is given by Equation 27 and

$$I_\gamma = \frac{\int r^2 (1+r^2)^{-3\beta} \, dr}{\int r^2 (1+r^2)^{-3\beta(\gamma+1+\alpha(\gamma-1))/(2+\alpha(\gamma-1))} \, dr} \qquad (38)$$

is a constant independent of the volume fraction distribution, $f$. At a given radius, the dependence of the total mass estimate on $f$ comes from Equation 23,

$$M_{\rm tot} \propto \frac{P}{<n>} \propto \kappa <n>^{\gamma-1} \propto \kappa A^{\gamma-1} . \qquad (39)$$

Hence from Equation 37 we see that the correction factor is equal to $C_T$, just as for the isothermal case.

Finally, Equation 37 can be substituted back into Equation 36 to eliminate $\kappa$, giving

$$<n> = C_n I_\gamma^{-\alpha/2} (1+r^2)^{-3\beta/(2+\alpha(\gamma-1))} , \qquad (40)$$

where once again the multiphase correction factor, $C_n$, is identical to the isothermal one given by Equation 30.

In summary, moving to a polytropic temperature law makes little difference to our conclusions. The baryon fraction within the core of the cluster can be lowered, but at the expense of a higher fraction in the cluster as a whole. The correction factors which need to be applied to the mass estimates when moving from a single-phase to a multiphase icm are identical to those for the isothermal case.

## 5 DISCUSSION

We have shown that, if the icm is multiphase throughout, as we know it to be within the cooling flow at its core, then estimates of the baryon fraction from X-ray observations need to be reduced by one quarter to one half. Taking a polytropic temperature distribution changes the estimate of the baryon fraction, making it lower in the cluster core (for $\gamma > 1$) but substantially higher at large radii, however the correction factor when moving from a single-phase to a multiphase icm is identical to the isothermal case.

In our model we have neglected a whole gamut of observational details. In particular we considered the bolometric flux emitted by the cooling gas. In reality this will be modified both by absorption, which tends to cut off the spectrum below a keV, and by the energy window of the detector. It is doubtful whether this would alter our conclusions greatly. The emission temperature in rich clusters is typically $3 \times 10^7$–$10^8$ K which gives a flat spectrum from bremsstrahlung emission across the energy band of most imaging detectors. In a multiphase icm there is a mixture of gas at both higher and lower temperatures which will contribute a lesser or greater fraction of its emission in the observed energy band, respectively. As all density phases contribute in roughly equal measure to the observed emission (the high-density phases are cooling most rapidly but occupy a much smaller volume) then the nett effect will be to leave the inferred bolometric emissivity almost unchanged.

A correction factor of 26 or even 49 percent is not enough to resolve the baryon catastrophe all on its own.

Some fraction of the Universe could be made up of hot dark matter (hdm) that does not cluster efficiently into cluster halos, but the hdm mass-fraction cannot be greater than about 30 percent which is still not sufficient to save the idea of an Einstein-de Sitter Universe.

Many simulations of cluster formation have been carried out with little evidence that baryons will concentrate preferentially in clusters. Both White et al. (1993) and Babul & Katz (1993) looked specifically for this effect but found a baryon enhancement of only about 20 percent in the cluster as a whole. It should be pointed out, however, that there could be a lot of missing physics in these simulations, for example baryons could be congregated by inhomogeneous ionizing radiation fields in the early universe.

Optical observations of cluster masses are prone to large errors, both through projection effects and because the three-dimensional shape of the velocity-dispersion tensor is unknown. However, X-ray mass determinations are much more reliable. Evrard (1994) suggests an accuracy of within ±50 percent, and Schindler (1995) an rms dispersion of 15 percent about the correct value, with no systematic under- or over-estimate.

One further way of increasing cluster mass estimates, more in the spirit of this paper, is to modify the equation of hydrostatic support for the icm by adding extra terms to the pressure (e.g. Loeb & Mao 1994). The extra contribution from magnetic fields and turbulence in equipartition could triple the inferred masses, thus negating the Baryon Catastrophe at a stroke. This hypothesis is supported by modelling of gravitational arcs which gives masses 2–2.5 times larger than those deduced from an unmodified equation of hydrostatic support (e.g. Miralda-Escudé & Babul 1995) although a correct modelling of shear may lower this estimate somewhat (Bartelmann, Steinmetz & Weiss 1995).

In a contrary vein, cold gas which has condensed from the icm will add to the observed baryon content of clusters and if it is coupled to the hot gas (a situation which has been termed 'mass loading') then it will lower the cluster mass estimate (by raising the mean density but not the pressure of the gas). We can only assume that this does not happen to a significant extent outside the cluster core, although there is evidence that it does do so within the cooling flow region (Daines, Fabian & Thomas 1994).

Our own personal view is that the intracluster medium is much more complex than most people have hitherto assumed and that there is sufficient uncertainty in its modelling so as to permit a critical density, Einstein-de Sitter Universe.

## ACKNOWLEDGMENTS

This paper was prepared using the facilities of the STARLINK minor node at Sussex. It was completed while PAT was holding a Nuffield Foundation Science Research Fellowship.

## REFERENCES

Allen S. W., Fabian A. C., Edge A. C., Böhringer H., White D. A., 1995, MNRAS, 275, 741
Bartelmann M., Steinmetz M., Weiss A., 1995, A&A, 297, 1
Böhringer H., 1994, in Durret F., Mazure A., J. Trân Thanh Vân, eds, Clusters of Galaxies. Editions Frontières, p. 139
Babul A., Katz N., 1993, ApJ, 406, L51
Buote D. A., Canizares C. R., 1996, ApJ, in press, astro-ph 9504049
Daines S. J., Fabian A. C., Thomas P. A., 1994, MNRAS, 268, 1060
Dekel A., 1994, ARA&A, 32, 371
Elbaz D., Arnaud M., Böhringer H., 1995, A&A, 293, 337
Evrard A. E., 1994, in Durret F., Mazure A., J. Trân Thanh Vân, eds, Clusters of Galaxies. Editions Frontières, p. 241
Fabian A. C., Hu E. M., Cowie L. L., Grindlay J., 1981, ApJ, 248, 47
Fabian A. C., 1991, MNRAS, 253, 29
Fabian A. C., 1994, ARA&A, 32, 277
Henriksen M. J., Mamon G. A., 1994, ApJ, 421, L63
Jones C., Forman W., 1984, ApJ, 276, 38
Loeb A., Mao S., 1994, ApJ, 435, L109
Miralda-Escudé J., Babul A., 1995, ApJ, 449, 18
Nulsen P. E. J., 1986, MNRAS, 221, 377
Sasselov D., Goldwirth D., 1995, ApJ, 444, L5
Schindler S., 1995, preprint, astro-ph 9503040
Thomas P. A., 1988, MNRAS, 235, 315
Thomas P. A., Fabian A. C., Nulsen P. E. J., 1987, MNRAS, 228, 973
Walker T. P., Steigman G., Schramm D. N., Olive K. A., Kang H. -S., 1991, ApJ, 376, 51
White D. A., Fabian A. C., 1995, MNRAS, 273, 72
White S. D. M., Navarro J. F., Evrard A. E., Frenk C. S., 1993, Nat, 366, 429